# Laser-Liquid Interaction in Laser-Induced Forward Transfer (LIFT) Printing: A Multiscale Perspective on Bubble Dynamics and Material Ejection

Shuqi Zhou, Abdol Hadi Mokarizadeh, and Ben Xu*
Department of Mechanical and Aerospace Engineering, University of Houston, Houston, Texas 77204, USA
*Address all correspondence to bxu12@Central.UH.EDU

## Abstract

Laser-induced forward transfer (LIFT) is a nozzle-free laser-assisted printing method in which a pulsed laser transfers material from a donor ribbon to a receiving substrate. Because the transferred material can be patterned directly from a computer-defined laser path without masks or nozzles, LIFT provides an advanced manufacturing route for spatially selective deposition of functional inks, nanoparticle suspensions, polymers, hydrogels, biological materials, and other difficult-to-nozzle formulations. The apparent simplicity of LIFT, however, conceals a strongly coupled laser-liquid interaction. Laser energy is absorbed within a confined donor architecture, converted into thermal, plasma, thermoelastic, vapor, or blister-like responses, and then transformed into bubble-mediated motion of the donor material. The cavitation bubble or expanding release volume provides the transient mechanical bridge between optical energy deposition and the hydrodynamic ejection process.

This chapter presents LIFT from a multiscale perspective centered on bubble dynamics and material ejection. It reviews major LIFT donor architectures first, including direct LIFT, absorbing-layer-assisted LIFT, dynamic release layers, blister-actuated LIFT, and sensitive-material transfer. Then, it examines how donor ribbon design, absorbing-layer properties, laser wavelength, pulse duration, spot size, fluence, material rheology, film thickness, receiver gap, and wettability control bubble inception, bubble growth, jet formation, droplet breakup, and final deposition. The chapter emphasizes that printing performance cannot be inferred from fluence alone; it depends on how laser energy is converted into a pressure impulse and how that impulse couples to a deformable material layer.

Modeling approaches are discussed as tools for connecting experimental observations across time and length scales. Modeling approaches are discussed as tools for connecting experimental observations across time and length scales, ranging from reduced-order estimates to interface-resolving simulations and data-driven process maps. As one illustrative mechanistic example, thermal-only, plasma-mediated, and coupled plasma-thermal-thermoelastic frameworks for early-stage bubble inception are briefly compared to show how different inception assumptions can provide initial conditions for downstream bubble growth and jetting models. This chapter concludes by identifying opportunities for bubble-aware donor design, time-resolved diagnostics, benchmark datasets, and predictive LIFT process maps based on intermediate bubble and jet observables.

## 1. Introduction: Laser-Liquid Interaction as the Hidden Core of LIFT

Laser-induced forward transfer (LIFT) is commonly described as a direct-write process: a laser pulse strikes a donor ribbon, propelling material toward a receiving substrate. This operational description is useful, but it hides the physical event that controls printing. Between the incident pulse and the final deposit, laser energy must be absorbed, converted into heat or pressure, localized into a transient expanding volume, coupled to a deformable donor material, focused into a jet or droplet, and finally deposited on a receiver [1-4]. In liquid-phase and soft-material LIFT, the cavitation bubble, vapor pocket, or blister-like release volume is therefore not a secondary by-product. It is the transient actuator that connects laser energy deposition to material ejection [5-7].

The attraction of LIFT lies in its ability to deposit materials that are difficult to print with nozzle-based printing. Because the material does not need to pass through a small nozzle, LIFT can handle a wider range of viscosities, particle loadings, biological sensitivities, and functional compositions than many conventional drop-on-demand methods [2, 4, 8]. This advantage also creates a challenge: there is no single universal ejection mechanism. A metal donor film, a nanoparticle ink, a hydrogel, a biological suspension, and a high-viscosity paste may all be printed by LIFT, but each responds differently to the laser-generated mechanical impulse.

Many experimental process windows are described in terms of incident fluence, spot size, donor thickness, and receiver gap [3-6]. These parameters are essential, but they are not the mechanism itself. Two donor systems can receive the same nominal fluence and produce different bubbles, jets, and deposits because their optical absorption, thermal confinement, stress confinement, material rheology, and boundary conditions differ [4-8]. Conversely, similar deposits may arise from distinct bubble histories. Therefore, a chapter focused solely on final printed shapes on the substrate would miss this underlying physics.

This chapter advances the view that laser–liquid interaction constitutes the hidden core of LIFT, with cavitation bubble dynamics acting as the fundamental mechanism that translates laser energy input into material ejection. Bubble inception determines where and how the mechanical actuator is born. Bubble growth determines the impulse applied to the donor layer. Bubble confinement determines directionality and free-surface focusing. Jet formation, breakup, and impact then determine the printed feature. The chapter follows this process from donor architecture and energy coupling through bubble dynamics, jetting, deposition, and modeling strategies.

The scope is intentionally broader than any single model. Early-stage plasma, thermal, and thermoelastic mechanisms are important in certain regimes, particularly under short-pulse or high-intensity conditions [9-11]. However, the chapter is not mainly about a single plasma-thermal model. Instead, that framework is used later as an illustrative example in a broader review of LIFT printing physics and multiscale modeling.

## 2. Fundamentals and Configurations of LIFT Printing

LIFT is a family of laser direct-write methods in which a pulsed laser releases material from a donor ribbon toward a receiver. The donor normally consists of a transparent carrier, an optically absorbing region, and a transferable material layer.

Depending on the donor architecture and material state, the laser pulse may vaporize or ablate a thin layer, form a cavitation bubble in a liquid film, deform a blister membrane, or drive a mechanically confined release event. These variants share the same practical goal: converting localized optical energy into a controlled impulse that transfers material without a nozzle [1, 12, 13].

In early direct LIFT, the laser interacted strongly with the material to be transferred. This approach established the basic capability to print metals and other functional materials, but it can expose the printed material to high thermal and mechanical loading [12-14]. Absorbing-layer-assisted LIFT separates optical absorption from the donor material. A metal, polymer, or other absorbing layer captures the pulse and indirectly drives transfer, which is useful when the functional material is weakly absorbing, highly viscous, particle-laden, or sensitive to direct irradiation [15-17].

Dynamic release layer and blister-actuated LIFT extend this separation further. In dynamic release layer concepts, a sacrificial layer decomposes or expands to generate the release impulse; in blister-actuated LIFT, a deformable membrane converts laser loading into a mechanical displacement that pushes the material forward [18-21]. These architectures are especially relevant when the process must reduce debris, limit direct contact between plasma and ink, or protect sensitive materials. They also highlight that the absorbing layer is not passive: its thickness, optical absorption, thermal response, adhesion, and mechanical compliance influence bubble formation, membrane motion, jet velocity, and repeatability.

Liquid-phase bubble-driven transfer is not treated here as a strictly separate donor architecture. Instead, it is a hydrodynamic actuation mode that often occurs within absorber-assisted or interface-absorbing donor designs, where localized energy deposition generates a cavitation or vapor bubble that drives free-surface deformation and jetting. Time-resolved imaging studies show that bubble size, shape, position, and lifetime are intermediate observables between laser input and the final printed feature [5-7]. This intermediate-bubble view is important because two experiments with similar fluence can produce different deposits if donor thickness, absorbing layer, receiver gap, or ink rheology changes the bubble-to-jet conversion.

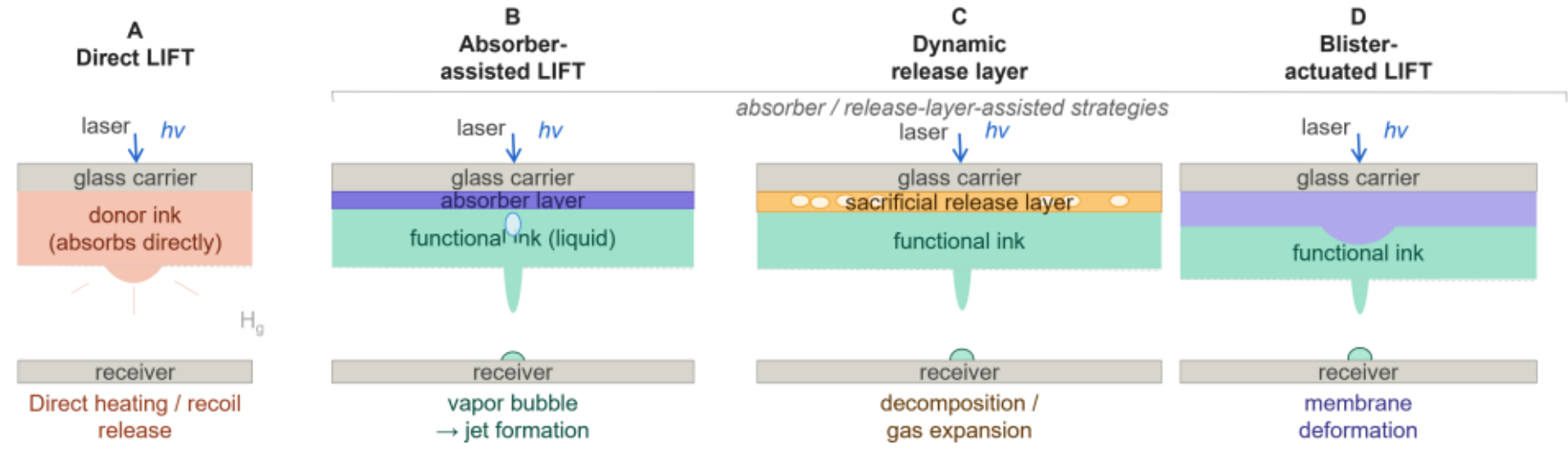

Figure 1. Representative LIFT donor architectures and actuation pathways. Direct LIFT relies on direct absorption by the transferable donor material, while absorber-assisted LIFT uses a separate absorbing layer to convert the laser pulse into a thermal, vapor, pressure, or bubble-mediated impulse. The dynamic release layer and blister-actuated LIFT are presented as specific absorber/release-layer-assisted strategies based on sacrificial layer decomposition and membrane deformation, respectively. Bubble-driven jetting is treated as a hydrodynamic actuation mode that may occur within absorber-assisted or interface-absorbing liquid donors rather than as a separate donor architecture. Bio-LIFT and laser-assisted bioprinting can use these pathways with hydrogels, bio-inks, or cell-laden materials.

Related laser direct-write methods, including Matrix-Assisted Pulsed Laser Evaporation Direct Write (MAPLE-DW) and laser-assisted bioprinting, broaden the material landscape of LIFT. They have been used for biological materials, hydrogels, particles, and functional inks where nozzle clogging, high viscosity, or cell viability can limit conventional printing [15, 22-25]. For this chapter, these variants are treated as distinct actuation pathways within the same physical problem: how laser energy converts into a localized mechanical impulse, how that impulse forms a bubble or deformation, and how the resulting motion ejects material. Representative donor architectures and actuation pathways are summarized in Figure 1, and their energy-absorbing components, actuation mechanisms, material classes, advantages, and limitations are compared in Table 1.

Table 1. LIFT configurations and actuation mechanisms. The table compares where laser energy is absorbed, how the release impulse is generated, representative material classes, advantages, and limitations for major LIFT donor architectures and application classes. Liquid-phase bubble-driven jetting is treated as a hydrodynamic mechanism that can occur within absorber-assisted liquid donor systems rather than as a separate donor architecture.

| Configuration | Energy-absorbing component | Actuation mechanism | Typical materials | Advantages | Limitations | Representative references |
|---|---|---|---|---|---|---|
| Direct LIFT | Printable donor itself | Local heating, vapor recoil, melting, ablation, or spallation drives release | Metals, absorbing films, some functional layers | Simple donor stack; strong coupling; useful for thin solid films | Direct thermal/mechanical exposure; possible debris or material damage | [12-14] |
| Absorber-assisted LIFT | Thin metal, polymer, or other absorbing layer | Absorber converts laser energy into heat, pressure, vapor, or bubble growth | Inks, nanoparticle suspensions, polymers, hydrogels | Separates optical absorption from printable material; broader material compatibility | Absorber design strongly affects impulse, contamination risk, and repeatability | [15-17, 30] |
| Dynamic release layer LIFT | Sacrificial release layer | Decomposition or expansion of release layer generates transfer impulse | Sensitive materials, biological fluids, functional inks | Reduces direct laser exposure; can moderate loading | Release chemistry and residue must be controlled | [18, 19] |
| Blister-actuated LIFT | Absorbing/deformable blister layer | Membrane deformation pushes donor material forward | Viscous inks, soft materials, microdevices | Mechanical actuation can be repeatable and less disruptive | More complex donor fabrication; membrane mechanics affect transfer | [20, 21, 33, 34] |
| Bio-LIFT / laser-assisted bioprinting | Absorbing/release layer or liquid donor region | Bubble, vapor, or blister impulse transfers biological ink | Hydrogels, cells, biomaterials | Nozzle-free sensitive-material patterning | Requires control of viability, shear, thermal exposure, and deposition quality | [15, 16, 24, 25, 30] |

## 3. Donor Ribbon Design and Laser Energy Coupling

The donor ribbon establishes the initial conditions for bubble-mediated LIFT. A typical donor includes a transparent carrier, an absorbing or transducing layer, a transferable material layer, and a receiver separated by a controlled gap. Each component affects how energy is distributed and how the resulting pressure couples to material motion.

The transparent carrier provides optical access and mechanical support. Its refractive index influences focusing and aberration. Its thermal properties affect heat removal from the absorbing layer. Its acoustic impedance influences pressure-wave reflection and transmission. In many liquid-film LIFT configurations, the carrier and absorber act as a relatively stiff upper boundary, while the lower free surface of the donor material is deformable. A bubble born near this stack therefore evolves in a confined geometry rather than in an infinite liquid.

In absorber-assisted LIFT, the absorbing layer is often a central design element for energy coupling. Its absorption coefficient, thickness, thermal conductivity, heat capacity, adhesion, mechanical compliance, and decomposition behavior determine how the laser pulse is converted into heat, pressure, vapor, plasma, or blister deformation. A strongly absorbing thin layer can localize energy near the donor interface. As an order-of-magnitude estimate, the thermal diffusion length during the pulse is

$$l_{th} \sim (\alpha \, \tau_L)^{\frac{1}{2}} \qquad \text{Eq. (1)}$$

where $l_{th}$ is the thermal diffusion length, $\alpha$ is thermal diffusivity, and $\tau_L$ is the laser pulse duration.

Shorter pulses and lower diffusivity therefore localize heating more strongly. A complementary stress-confinement estimate is

$$t_a \sim \frac{L}{c} \qquad \text{Eq. (2)}$$

where $t_a$ is the acoustic relaxation time, $L$ is the characteristic heated or absorbing length scale, and $c$ is the speed of sound.

Pressure transients can become important when energy is deposited before the heated region relaxes acoustically [26, 27]. If the layer decomposes or delaminates, gas formation can contribute to a blister or expanding cavity. If the absorber remains intact, thermoelastic deformation may dominate.

The donor thickness controls both the available transfer volume and the distance the bubble or blister impulse must travel before reaching the free surface. If the transferable layer is too thick relative to the bubble size or pressure impulse, deformation may remain mostly internal, and no jet forms. If it is too thin, the expanding volume may rupture the layer, produce spray, or generate unstable ejection. An intermediate range can support coherent free-surface focusing and droplet formation. Thickness also affects jet diameter, heat capacity, hydrodynamic resistance, and sensitivity to local coating nonuniformity [6, 7, 16].

Laser parameters provide external control over the actuator. Wavelength selects which layer absorbs energy. Pulse duration determines whether the response is closer to quasi-thermal heating, stress-confined thermoelastic loading, optical breakdown, or ablation-driven release. Spot size sets the lateral scale of the bubble footprint and therefore affects jet diameter and deposited feature size. Fluence or pulse energy controls the available energy, but the useful ejection work is only a fraction of the absorbed energy. Some energy remains as heat, some is lost to acoustic radiation, some drives phase change or chemical decomposition, and some is dissipated by viscosity.

Material properties determine how the donor layer interprets the bubble impulse. Viscosity damps rapid deformation. Surface tension resists the formation of interfaces and drives breakup. Density controls inertia. Elasticity can resist jet thinning or promote filament stretching. Particle loading can change viscosity, optical scattering, drying behavior, and final functional properties. Biological materials introduce additional constraints on temperature, pressure, shear, and chemical exposure.

The receiver gap and receiver surface complete the energy-to-deposit chain. A short gap can allow a jet to contact the receiver before capillary breakup. A larger gap can allow satellite droplets to evaporate, deviate from their trajectories, or break up before impact. Receiver wetting, roughness, temperature, and compliance influence spreading, splashing, immobilization, drying, and final printing patterns [7, 28, 29].

The absorbing layer and donor material should be treated as a coupled pair rather than as independent design choices. A highly absorbing layer beneath a low-viscosity liquid may create a strong initial deformation and a satellite-prone jet. The same absorber beneath a viscous or viscoelastic ink may be necessary to reach the transfer threshold. A mechanically compliant release layer may reduce shock loading of a biological material but may also broaden the impulse and reduce jet speed. These tradeoffs make LIFT design an exercise in matching source strength, source duration, and material response [30].

Laser wavelength, beam quality, and pulse-to-pulse repeatability also influence energy coupling. Wavelength controls which layer absorbs and how deeply energy penetrates. A donor stack designed for ultraviolet absorption may behave very differently under visible or near-infrared irradiation if the absorber, solvent, or particles have different optical responses. Beam profile determines whether the source is centrally peaked, ring-like, or approximately top-hat. Pulse-to-pulse fluctuations can shift a process near the threshold between no-transfer, stable-transfer, and spray regimes. For high-throughput printing, these variations become as important as average fluence [13, 31, 32].

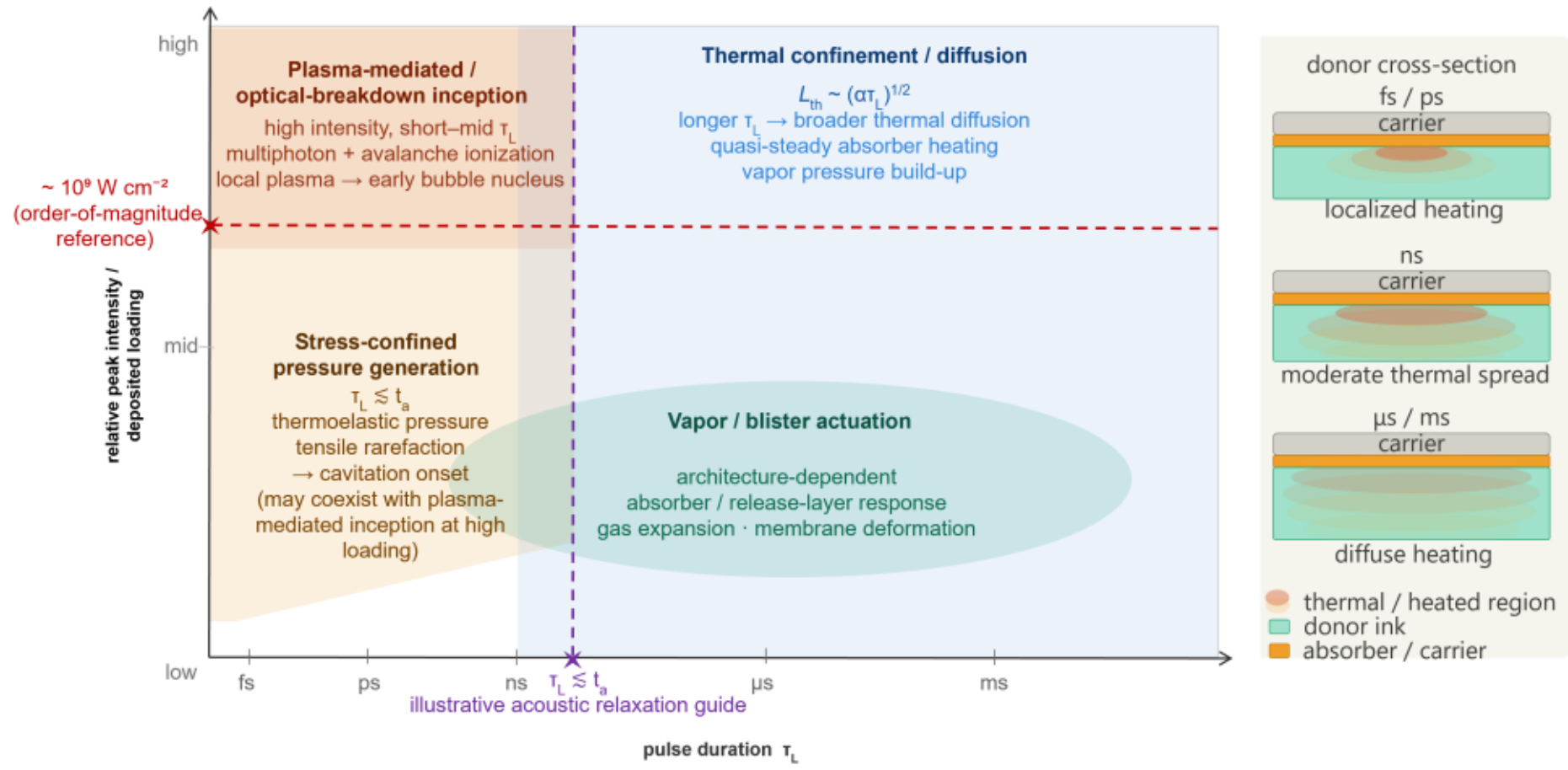

Figure 2. Schematic donor energy-coupling and confinement map for LIFT. The horizontal axis represents laser pulse duration, and the vertical axis represents relative peak intensity or deposited loading. Short-pulse loading can generate stress-confined pressure transients when $\tau_L \lesssim t_a$, where $t_a = L/c$ is the acoustic relaxation time, $L$ is the characteristic heated or absorbing length scale, and $c$ is the speed of sound. Stress confinement is treated here as a time-scale condition for rapid pressure generation rather than as an independent bubble-inception mechanism, and it may coexist with plasma-mediated or optical-breakdown-assisted inception under high-loading short-pulse conditions. As pulse duration increases, thermal diffusion becomes more important, and the characteristic thermal diffusion length scales as $l_{th} \sim (\alpha\tau_L)^{1/2}$, leading to broader heating and stronger contributions from absorber heating and vapor pressure. Vapor or blister actuation is shown as architecture-dependent because it also depends on absorber thickness, release-layer chemistry, adhesion, gas generation, and membrane compliance. The horizontal dashed line is only an order-of-magnitude reference, and the vertical dashed guide marks an illustrative location rather than a universal boundary.

A final donor-design issue is practical repeatability. In repeated printing, changes in absorber response, donor-layer uniformity, or blister dynamics may alter subsequent transfer events even when the nominal laser parameters remain fixed. These effects can affect process repeatability and remain an area where more systematic study would be useful. Practical process windows therefore require not only one-shot transfer thresholds but also repeatability and process-control metrics over many pulses and positions [21, 33, 34].

Donor design should therefore be understood as actuator design. Effective donor design does not simply maximize absorption. It shapes the spatial footprint, rise time, duration, and directionality of the pressure source so that the bubble or blister trajectory matches the material and receiver. These confinements and energy-coupling ideas are organized schematically in Figure 2.

## 4. Cavitation Bubble Dynamics in LIFT

Cavitation bubble dynamics provide the mechanical bridge between laser energy deposition and liquid ejection. In liquid-phase LIFT, the laser does not simply push a droplet from the donor. Instead, absorbed energy produces a localized pressure and temperature disturbance that nucleates a bubble. The bubble then expands, deforming the liquid film and focusing momentum toward the free surface. Its growth curve, shape, position, and collapse behavior help determine whether the outcome is clean jetting, broad ejection, splashing, or no transfer.

Bubble inception may occur through several pathways. Local heating can superheat the liquid and promote vapor nucleation, particularly near absorbing surfaces or heterogeneous nucleation sites [35-37]. Under short-pulse or high-intensity conditions, optical breakdown, plasma-mediated heating, and rapid pressure-wave generation may also contribute [9, 10]. In LIFT, the relevant question is not which mechanism is universal, but which mechanism supplies the initial bubble size, pressure, and geometry for the donor architecture and pulse regime being used.

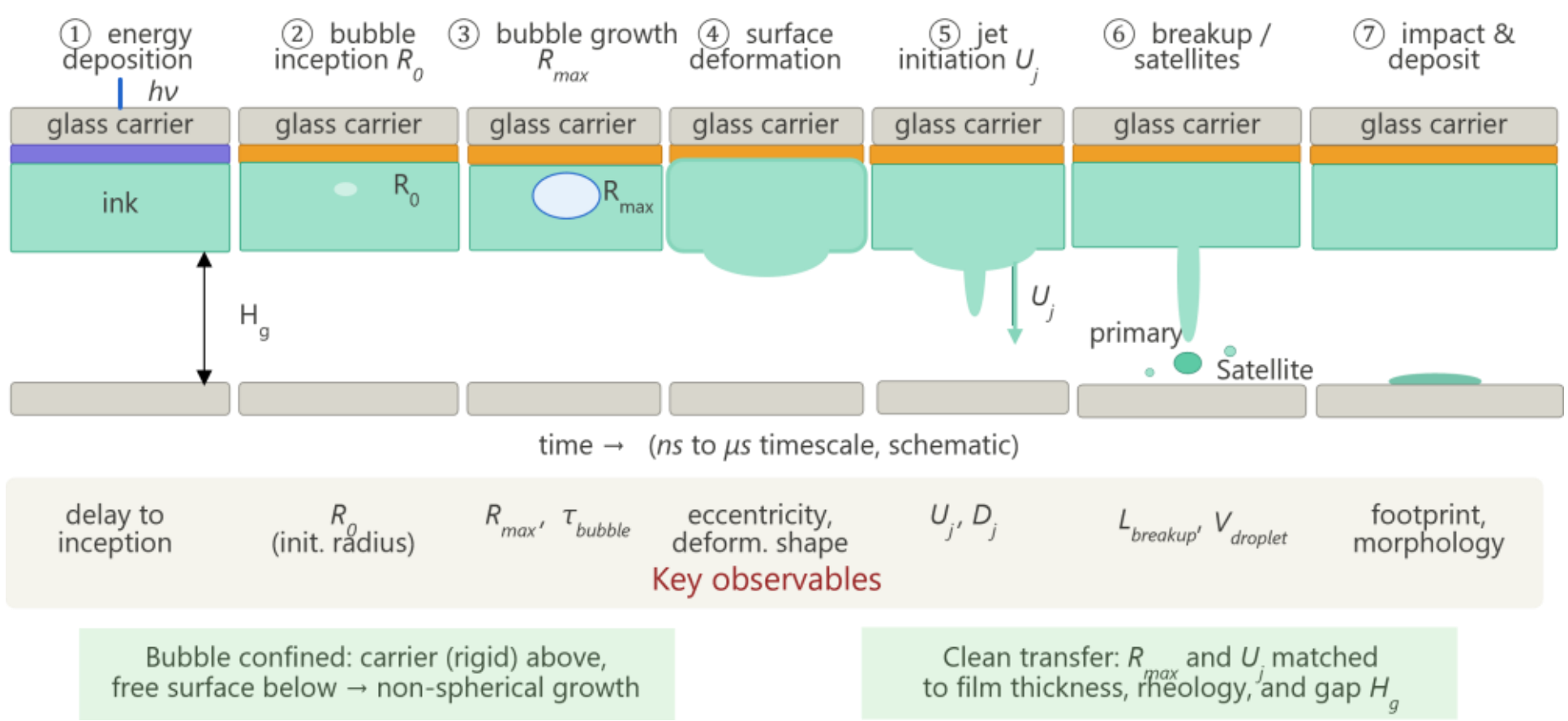

Figure 3. Bubble-mediated jet formation in a forward-transfer geometry. After localized energy deposition, a bubble, vapor pocket, or blister-like expansion forms near the donor interface, grows under geometric confinement, deforms the free surface, initiates a jet, and drives stretching, breakup, impact, and deposition. Key intermediate observables include the initial bubble size $R_0$, maximum bubble response $R_{max}$, bubble lifetime $\tau_{bubble}$, jet speed $U_j$, jet diameter $D_j$, receiver gap $H_g$, breakup state, and final deposit pattern. These observables provide validation targets between the laser input and the printing outcome.

After inception, the early expansion is usually inertia-dominated. The bubble pressure exceeds the surrounding pressure and drives liquid outward; viscosity and surface tension are initially secondary but become important as the bubble slows, approaches its maximum radius, or interacts with nearby interfaces [38, 39]. Classical bubble theory is useful for interpreting growth time and energy scaling, but LIFT bubbles are rarely spherical or unbounded. They form close to a carrier, absorbing layer, deformable donor surface, free surface, and sometimes a nearby receiver, so confinement and asymmetry are central rather than secondary [40, 41]. The confined bubble-to-jet sequence and the associated intermediate observables are summarized in Figure 3.

The bubble-to-jet transition occurs when expansion displaces the donor film and amplifies a free-surface protrusion into a jet. Bubble radius and lifetime provide a useful measure of impulse, while bubble position and shape govern how that impulse is directed. A bubble too weak or too far from the free surface may only deform the film; a stronger or better-positioned bubble can initiate a jet; an excessive impulse can rupture the film, generate satellites, or splash the receiver [5, 6]. Collapse and rebound may also affect the final outcome, especially in confined geometries where reflected pressure waves and nearby boundaries redirect motion [10, 42].

Several experimental observables are therefore more informative than fluence alone: delay to visible bubble formation, initial and maximum radius, growth and collapse time, bubble eccentricity, jet initiation time, jet speed, jet diameter, breakup length, droplet volume, and deposition footprint. These observables allow experiments and models to meet at intermediate stages rather than comparing only the laser input to the final deposit. They also help identify whether a poor feature originates from energy absorption, bubble growth, free-surface deformation, rheology, or impact.

This bubble-centered interpretation clarifies why donor boundaries matter. The same bubble in an infinite liquid would expand and collapse differently from a bubble near a wall or free surface. In LIFT, the carrier constrains one side, the free surface provides a deformable boundary, and the receiver may alter the pressure field or intercept the jet. Confinement can enhance directionality, but it also narrows the window between clean transfer and disruptive ejection [43-45].

## 5. Jet Formation, Material Ejection, and Deposition

Jet formation is the stage at which bubble dynamics become printing performance. As the bubble or blister expands, it displaces the donor material and deforms the free surface. If the pressure impulse, film thickness, and material properties are properly matched, the deformation focuses into a forward jet. If the impulse is too weak, the surface may bulge and retract; if it is too strong or poorly distributed, the donor may rupture, atomize, or produce spray.

Jet initiation depends on bubble size, growth rate, position, and footprint. A rapidly expanding bubble near the donor interface can accelerate the free surface and generate a narrow jet, while a broader or slower bubble may produce a wider protrusion with lower velocity. Bubble asymmetry can tilt the jet or produce uneven breakup. These effects explain why stable printing cannot be predicted from fluence alone: the same pulse energy may produce different jets if film thickness, absorber response, or material rheology changes.

Jet speed and diameter determine flight and impact conditions. A fast, slender jet may reach the receiver before capillary breakup if the gap is short; a longer path allows capillary instabilities to grow, producing primary droplets and satellites. Jet breakup can be interpreted using standard dimensionless groups such as the Weber number, which compares inertial forcing with capillary resistance, and the Ohnesorge number, which compares viscous damping with inertial-capillary motion. In scaling form, $We \sim \rho U^2 D/\sigma$ and $Oh \sim \mu/(\rho\sigma D)^{1/2}$, where $\rho, U, D, \sigma$, and $\mu$ denote density, characteristic jet velocity, jet or droplet diameter, surface tension, and dynamic viscosity, respectively. These groups help interpret breakup and satellite trends, although direct bubble and jet observables remain necessary in transient LIFT geometries [50].

Receiver impact transforms the ejected material into a printed feature. Impact depends on velocity, volume, viscosity, surface tension, substrate wettability, roughness, and temperature, with outcomes including compact droplets, spread disks, splashes, elongated deposits, rings, and satellite patterns [29, 51]. The receiver can preserve or erase the memory of the jet: a coherent jet may form a compact dot on a wettable substrate, while the same jet can recoil, bead, or splash on a different surface.

Deposition pattern is therefore a record of earlier dynamics, but not a unique fingerprint. A central spot with satellites may reflect jet breakup, secondary ejection, or impact splashing. Irregular debris may indicate donor damage, excessive energy, or poor wetting. Mechanistic interpretation improves when final deposits are paired with time-resolved imaging of bubbles and jets.

Material integrity is a parallel performance metric. Conductive inks require continuous particle networks after drying or sintering; optical or electronic materials may require uniform thickness and low contamination; hydrogels and biological inks may require preserved network structure, viability, phenotype, and spatial organization [4, 5, 15, 16, 24]. These application-level constraints feed back into acceptable bubble and jet regimes. A violent jet may print a spot, but it may not print a useful device or viable biological pattern.

## 6. Governing Parameters and Bubble-Mediated Process Windows

LIFT performance is governed by the coupled parameters of the laser, donor, material, and receiver. Fluence is often the most visible control variable, but it is not a complete description because the same fluence can correspond to different pulse energies, spot sizes, pulse durations, absorption depths, and pressure histories. In a bubble-mediated view, the useful process window is better described by how those inputs determine bubble inception, maximum radius, jet initiation, jet stability, and deposition pattern [5, 52, 53].

Laser pulse energy, spot size, wavelength, and pulse duration should be considered as a coupled loading condition rather than separate knobs. Pulse energy and spot size set the lateral footprint and intensity of the source; wavelength determines which layer absorbs; and pulse duration controls how much heat or pressure can relax during loading. In process-window terms, these variables shift the system among no transfer, coherent jetting, satellite formation, and spray or damage [10, 42, 54].

Absorbing-layer properties govern the conversion from optical energy to mechanical actuation, but their practical effect depends on the surrounding donor and receiver conditions. Optical absorption depth, heat capacity, thermal conductivity, decomposition or melting behavior, adhesion, roughness, and compliance can change pressure history and bubble nucleation. For blister-based or dynamic-release designs, repeatability may also depend on how the release layer changes over repeated pulses, which remains a process-control issue rather than a fully resolved design rule [31, 32, 55].

Material properties determine how the same bubble impulse is expressed as motion. A viscous or viscoelastic donor may require a stronger impulse or shorter receiver gap, whereas a low-viscosity donor may require reduced impulse or better receiver control to avoid satellites and splashing. Particle, cell, or polymer loading adds constraints on jet stability, drying, material function, and biological integrity [46-48].

The receiver gap and surface condition complete the parameter set. Gap height affects whether the jet impacts as a continuous filament or a separated droplet and how much time it has to stretch, break, or deviate. Receiver wettability and surface energy control spreading and pinning after impact. In practice, a robust regime is defined by ranges of laser input, donor architecture, film thickness, material rheology, and receiver conditions that jointly produce repeatable bubble and jet observables.

Conceptual regime maps can organize these dependencies. One axis may represent laser energy or bubble impulse, while another represents material resistance such as viscosity or Ohnesorge number; additional annotations can mark donor thickness, spot size, and gap. Regions might include no transfer, clean jetting, satellite-forming jetting, splashing, and excessive donor disruption. Such maps should be treated as material- and architecture-specific, but they help shift optimization away from fluence-only thresholds toward bubble-aware printing design.

Figure 4 provides a conceptual process-window map based on bubble impulse and material resistance. The corresponding controllable parameters and their expected effects on bubble formation, jetting, and deposition are summarized in Table 2.

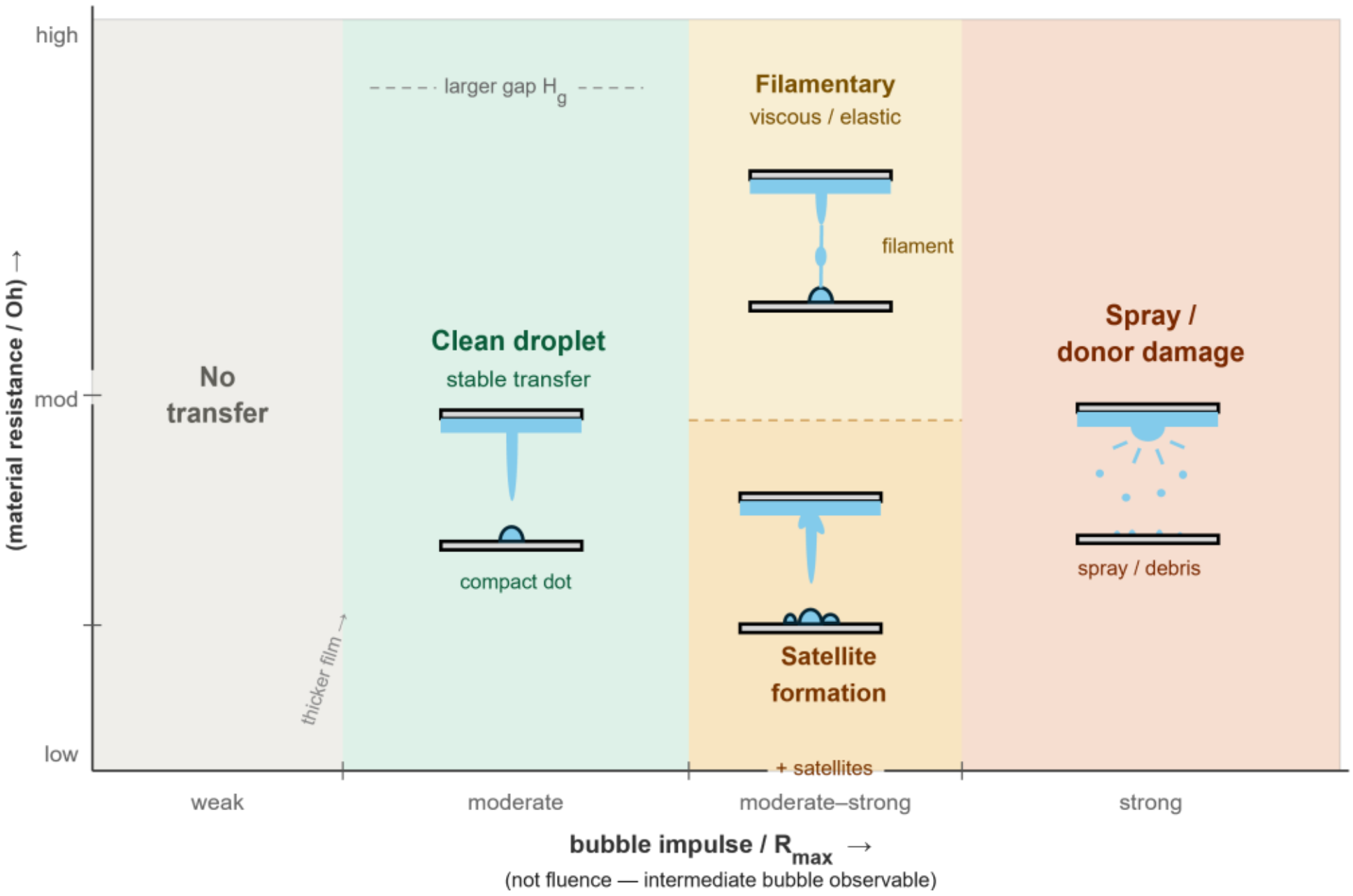

Figure 4. Schematic bubble-aware process-window map for LIFT material ejection. The horizontal axis represents effective bubble impulse or bubble-growth response rather than incident fluence alone, while the vertical axis represents material resistance associated with viscosity, elasticity, surface tension, and rheological damping. Conceptual regimes include no transfer, clean droplet transfer, filamentary transfer, satellite formation, and spray or donor damage. Laser, donor, material, and receiver settings influence the final outcome through intermediate bubble and jet observables such as maximum bubble response, jet speed, breakup state, receiver-gap effects, and deposit pattern. The boundaries are qualitative and depend on donor architecture, film thickness, absorber response, receiver gap, and material properties.

Table 2. Key physical parameters controlling bubble-mediated LIFT. The table links controllable laser, donor, material, and receiver variables to their primary physical roles, bubble/jet consequences, and deposition implications. Entries are intended as regime-dependent design guidance rather than universal rules.

| Parameter group | Parameter | Primary physical role | Bubble/jet consequence | Deposition implication | Representative references |
|---|---|---|---|---|---|
| Laser input | Wavelength | Selects absorbing layer and optical penetration | Changes source location and energy localization | Affects material exposure and feature resolution | [1-4] |
| Laser input | Pulse duration | Controls thermal and stress confinement | Changes pressure impulse, heating, and inception pathway | Influences jet violence, damage risk, and repeatability | [10, 26, 27, 54] |
| Laser input | Fluence / pulse energy | Sets available actuation energy | Changes bubble size, jet speed, and transfer threshold | Too low: no transfer; too high: spray or damage | [5, 52, 53] |
| Beam geometry | Spot size / profile | Sets lateral source scale | Changes bubble footprint and jet diameter | Controls feature size and satellite tendency | [31, 32] |

| Donor architecture | Absorbing/release layer | Converts optical energy into mechanical loading | Controls pressure history and bubble/blister formation | Determines transfer efficiency and material protection | [18-21, 30] |
|---|---|---|---|---|---|
| Film geometry | Donor film thickness | Sets bubble-to-surface distance and material volume | Controls surface focusing, jet stability, and transferred volume | Too thin may rupture; too thick may suppress jetting | [6, 7, 16] |
| Material properties | Viscosity, surface tension, elasticity | Balance inertia, damping, capillarity, and filament stretching | Controls jet breakup, satellites, and detachment | Determines clean droplet, filament, or failed transfer | [46-50] |
| Receiver conditions | Gap and wettability | Controls flight time, impact, spreading, and pinning | Changes breakup before impact and post-impact flow | Determines footprint, splashing, alignment, and final pattern | [28, 29, 51] |

## 7. Modeling Strategies for Bubble-Mediated LIFT

Modeling LIFT requires multiple levels of description because no single model can resolve the entire chain from optical absorption to the final deposit with equal fidelity. The appropriate model depends on the question being asked: threshold prediction, bubble growth, jet formation, deposition pattern, material integrity, or process optimization. A useful modeling hierarchy, therefore, moves from low-cost scaling estimates to bubble-dynamics models, interface-resolving simulations, and reduced descriptions for design.

Scaling and energy-balance models provide the first level. They estimate absorption depth, thermal diffusion length, acoustic relaxation time, capillary time, viscous time, bubble growth time, and jet flight time. These estimates help identify which measurements and simulations are worth pursuing. For example, if acoustic relaxation is faster than laser loading, a stress-confined model may not be necessary; if capillary breakup is slower than jet flight, breakup may occur after impact rather than during flight. Scaling models organize regimes and guide experiments, but they generally cannot predict detailed bubble shape or breakup.

Rayleigh-Plesset-type models describe the evolution of an equivalent spherical bubble radius by balancing liquid inertia, internal pressure, ambient pressure, surface tension, and viscosity [38, 39, 56]. A common reduced-order form is

$$\rho(RR'' + \frac{3}{2}(R')^2) = p_B(t) - p_\infty - \frac{2\sigma}{R} - \frac{4\mu R'}{R} \qquad \text{Eq. (3)}$$

where $R$ is the equivalent bubble radius, $\rho$ is the liquid density, $p_B(t)$ is the bubble pressure, $p_\infty$ is the ambient pressure, $\sigma$ is the surface tension, and $\mu$ is the dynamic viscosity.

CFD methods, including volume-of-fluid, level-set, and related interface-tracking approaches, are needed when free-surface deformation, jetting, breakup, and impact must be resolved [57-59]. CFD can include donor film thickness, receiver gap, surface tension, viscosity, contact angle, and boundary effects [60]. It can reveal velocity and pressure fields that are not experimentally accessible. However, CFD requires physically meaningful initial conditions and material models. If the bubble pressure or initial cavity is imposed arbitrarily, the simulation may reproduce a jet without explaining how the laser created it.

Thermal-fluid and phase-change models connect heat transfer to vapor generation, residual thermal energy, and bubble pressure. They are useful when absorber heating, solvent vaporization, or material damage are important. Plasma and thermoelastic source models are relevant under high-intensity or short-pulse conditions, where optical breakdown, rapid local heating, and stress confinement may contribute to the inception of cavitation [26, 61-63]. These models should be applied in a regime-dependent manner, rather than assumed for all LIFT configurations.

Reduced-order and data-driven models can support process optimization when sufficient experimental data are available. A useful reduced model should rely on intermediate variables such as bubble radius, lifetime, jet speed, breakup state, and deposit size; models trained only on fluence and final spot size may not transfer when donor architecture or material changes [64].

Model validation should use intermediate dynamics, not only final deposits. Useful validation targets include plasma or absorption shape, first visible bubble radius, radius-time curve, jet initiation time, jet velocity, breakup length, impact pattern, and deposit profile. This validation hierarchy connects early laser-liquid interaction to printing performance.

Energy-balance checks and CFD-boundary choices should be matched to the target observable. A measured bubble should be consistent with the absorbed energy after accounting for thermal losses, compressibility, and confinement; if not, the assumed absorption, pressure, or geometry should be revisited. Jet speed and breakup may be captured with a multiphase model and prescribed bubble pressure, while inception thresholds require optical absorption, heat transfer, phase change, or pressure-wave generation, and deposition pattern may require wetting, evaporation, solidification, or particle transport.

As LIFT modeling moves toward process design, uncertainty in absorber thickness, beam profile, donor-film uniformity, and material properties should also be considered. Although LIFT-specific uncertainty quantification studies remain limited, uncertainty-aware modeling strategies developed in additive manufacturing offer a useful direction for robust process optimization [65].

## 8. Atomistic and Electronic-Structure Inputs for Source-Term Modeling

Continuum LIFT models often prescribe absorbed energy, pressure, temperature, optical absorption, or initial bubble conditions. Atomistic and electronic-scale methods extend the hierarchy downward by supplying source-term and inception inputs. They are useful when nanometer-scale or femtosecond-to-picosecond response times matter, but they do not replace Rayleigh-Plesset-type, CFD/VOF, level-set, or deposition models for visible bubble growth, jetting, and printed pattern.

Classical molecular dynamics can resolve vapor-bubble nucleation in superheated or stretched liquids and estimate critical nucleus size, nucleation rate, local pressure, vapor density, and interfacial constraints [68]. It can also distinguish homogeneous nucleation from heterogeneous nucleation near surfaces, inclusions, nanoparticles, or donor interfaces [69, 70]. These results can constrain initial cavity, pressure, or threshold assumptions used by larger-scale LIFT models.

Atomistic-continuum and two-temperature molecular dynamics are useful for short-pulse laser loading of absorbing layers, metal films, or heated donor interfaces. Such models can describe electron-phonon coupling, melting, spallation, phase explosion, explosive boiling, or vapor-layer formation near heated boundaries [71].

They constrain early energy partitioning and phase change, while continuum models remain necessary for later confined-bubble and jet motion.

Density functional theory (DFT), ab initio molecular dynamics (AIMD), and real-time time-dependent DFT (RT-TDDFT) provide complementary source-term inputs that include electronic effects. DFT and AIMD can provide near-equilibrium values for optical constants, absorption coefficients, equations of state, thermal and elastic constants, interfacial energies, and decomposition pathways. RT-TDDFT is relevant to ultrafast electronic excitation, nonlinear absorption, ionization, optical breakdown, and plasma-mediated inception under high-field conditions [72, 73].

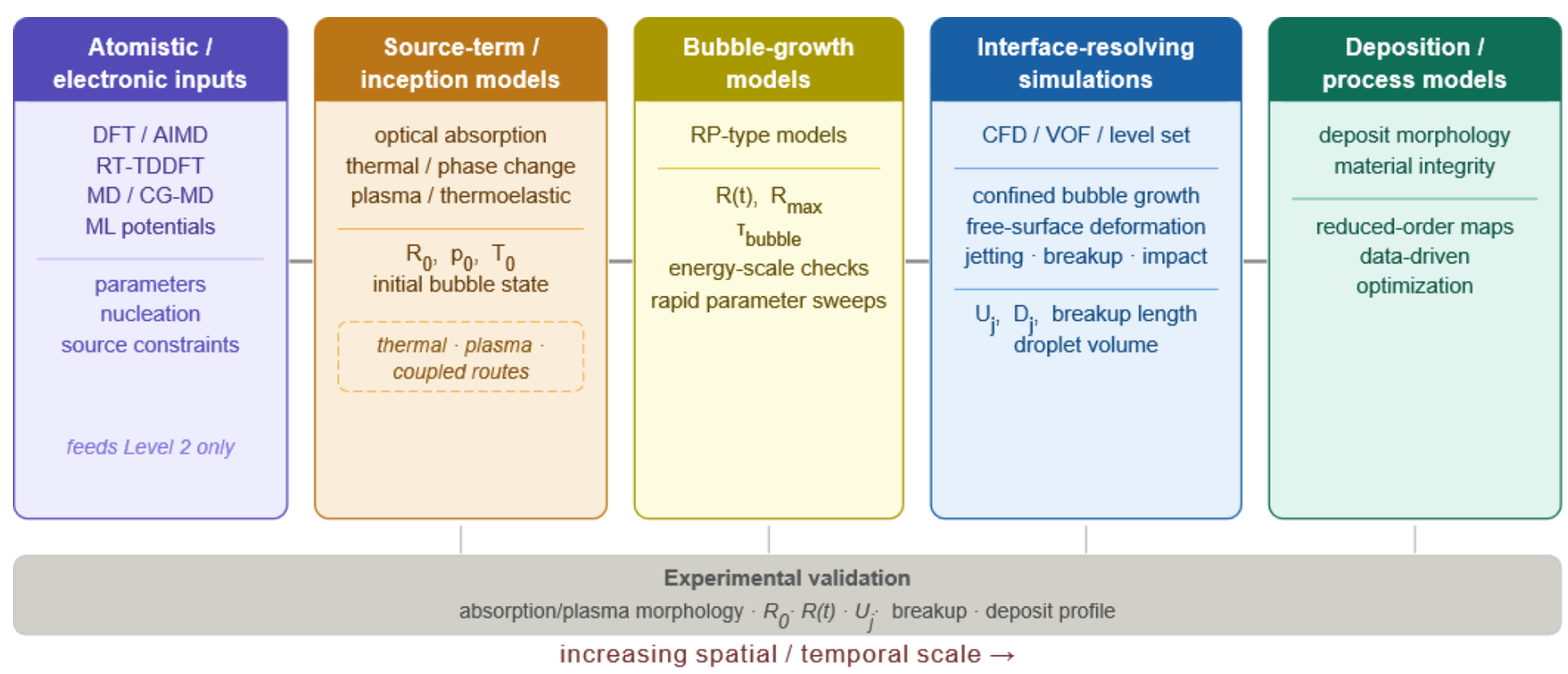

Figure 5. Multiscale modeling hierarchy for bubble-mediated LIFT. Atomistic and electronic-scale methods, including DFT, AIMD, RT-TDDFT, molecular dynamics, coarse-grained MD, and machine-learning interatomic potentials, provide material parameters, absorption and ionization physics, nucleation behavior, interfacial properties, and source-term constraints. These lower-scale inputs feed source-term and inception models rather than directly predicting visible jetting or deposits. Source-term models then connect to Rayleigh-Plesset-type bubble-growth models, CFD/VOF or level-set simulations of confined bubble growth and jetting, and deposition or process models. Experimental observables such as absorption, plasma shape, $R_0$, $R(t)$, $U_j$, breakup state, and deposit profile serve as validation checkpoints throughout the hierarchy.

Coarse-grained MD aims to bridge the scale gap between atomistic and continuum models, as it extends atomistic reasoning to membranes, hydrogels, polymer networks, and other soft or biological materials under rapid strain, pressure transients, or bubble-collapse loading [74]. Machine-learning interatomic potentials trained on first-principles data may extend atomistic accuracy toward larger systems and longer times, but uncertainty and training-domain limits must be tracked [75]. Used this way, lower-scale methods feed source-term and inception models rather than directly predicting full LIFT transfer. The resulting modeling hierarchy, including atomistic source-term support and continuum jetting models, is summarized in Figure 5.

The modeling approaches discussed above are summarized in Table 3, including their main capabilities, required inputs, validation observables, limitations, and suitable uses.

Table 3. Modeling approaches for bubble-mediated LIFT and suitable regimes, including atomistic/electronic-scale inputs, source-term descriptions, bubble models, interface-resolving simulations, deposition/material models, and reduced-order maps.

| Modeling approach | Captures well | Required inputs | Best validation observables | Main limitations | Suitable use | Representative references |
| --- | --- | --- | --- | --- | --- | --- |

| Scaling / energy balance | Thresholds, maximum bubble scale, rough jet trends | Absorbed energy, material properties, geometry | Transfer threshold, maximum bubble radius, jet onset | Needs empirical efficiency; limited geometry detail | Rapid estimates and process-map structure | [26, 27, 38, 39] |
|---|---|---|---|---|---|---|
| Source-term models | Optical absorption, heating, phase change, pressure generation | Optical constants, pulse shape, absorber properties, thermal data | Absorption depth, early pressure, plasma or heated-zone shape | Often needs assumptions about energy partition | Initial conditions for bubble models | [9, 10, 26, 61-63] |
| Illustrative inception routes | Thermal-only, plasma-mediated, and coupled plasma-thermal-thermoelastic pathways | Pulse regime, absorption geometry, ionization/heating assumptions | Delay time, first bubble size, early bubble shape | Regime-dependent; not universal | Mechanistic interpretation of early bubble inception | [35-37, 66] |
| Rayleigh-Plesset-type models | Equivalent-radius bubble growth and collapse | Initial radius/pressure, fluid properties, ambient pressure | Radius-time curve, maximum radius, lifetime | Assumes near-spherical bubble and simplified boundaries | Intermediate bubble scaling and calibration | [38, 39, 56] |
| CFD / VOF / level-set | Confined bubble growth, free-surface deformation, jetting, breakup | Geometry, initial bubble/pressure field, fluid properties, boundary conditions | Interface shape, jet speed, breakup state, droplet volume | Computational cost; sensitive to initial/boundary conditions | Detailed LIFT jet and deposition simulations | [57-60] |
| Reduced-order / data-driven models | Process maps, optimization, uncertainty, parameter screening | Curated datasets with laser, donor, bubble, jet, and deposit variables | Prediction accuracy across bubble, jet, and deposit observables | Limited extrapolation without physics-informed inputs | Design exploration and robust process control | [64, 65] |
| Atomistic / electronic-scale methods: DFT, RT-TDDFT, MD, coarse-grained MD, ML potentials | Electronic excitation, optical/material parameters, nucleation, early phase change, interfacial properties, soft-material integrity | Material composition, interatomic potentials or first-principles setup, excitation conditions, thermodynamic state | Ionization threshold, absorption response, nucleation tendency, interfacial properties, early inception constraints, material damage indicators | Nanometer/femtosecond-to-picosecond scale, high computational cost, not direct jet/deposit prediction | Source-term parameterization, inception-level physics, material-property input, scale-bridging support | [68-75] |

## 9. Illustrative Mechanistic Framework: Early-Stage Bubble Inception

As an illustrative example of early-stage bubble inception modeling, one can compare thermal-only, plasma-mediated, and coupled plasma-thermal-thermoelastic descriptions. This example is included not to claim that optical breakdown governs all LIFT processes, but to demonstrate how early-stage source physics can be translated into physically meaningful initial conditions for bubble-growth and jetting models. The comparison is not intended to represent all LIFT regimes. Rather, it shows how different physical assumptions about the first nanoseconds of laser-liquid interaction can, under certain conditions, lead to different initial conditions for downstream RP or CFD models [66].

In a thermal-only description, absorbed laser energy is treated primarily as a heat source, and the resulting superheated liquid state can be interpreted using vapor-bubble nucleation concepts such as vapor pressure, critical radius, nucleation rate, and nucleation probability [35, 36]. This route can provide a useful limiting description when heating and phase-transition kinetics dominate, such as in some absorber-assisted or longer-pulse regimes. Its limitation is that it may be incomplete when the observed inception occurs on a timescale comparable to stress relaxation or optical breakdown.

In a plasma-mediated description, a high-intensity pulse can generate seed electrons through multiphoton ionization and amplify them through avalanche ionization. Free electrons absorb energy from the laser field, creating a localized plasma absorption region [9, 10]. If the deposited energy is spatially elongated or asymmetric, the earliest cavity may inherit aspects of that geometry [42]. This route can help interpret non-spherical inception under optical-breakdown conditions, but by itself it may not capture longer-term thermally sustained bubble growth.

A coupled plasma-thermal-thermoelastic description treats localized energy deposition as a common source for both mechanical and thermal responses. Rapid deposition under stress confinement can generate a thermoelastic pressure field, while acoustic relaxation may create tensile rarefaction that can contribute to cavitation under certain conditions [26, 61, 62]. Residual heat can then contribute to vapor pressure and, later, to bubble growth. Figure 6 presents a representative validation case from our recent laser-induced cavitation study [66], in which thermal-only, plasma-mediated, and coupled descriptions are compared against experimentally observed bubble-radius evolution at selected pulse energies. This framework can provide a useful bridge between optical breakdown, early pressure transients, and continuum bubble dynamics, but it should be used carefully and only in regimes where the pulse and intensity justify such physics.

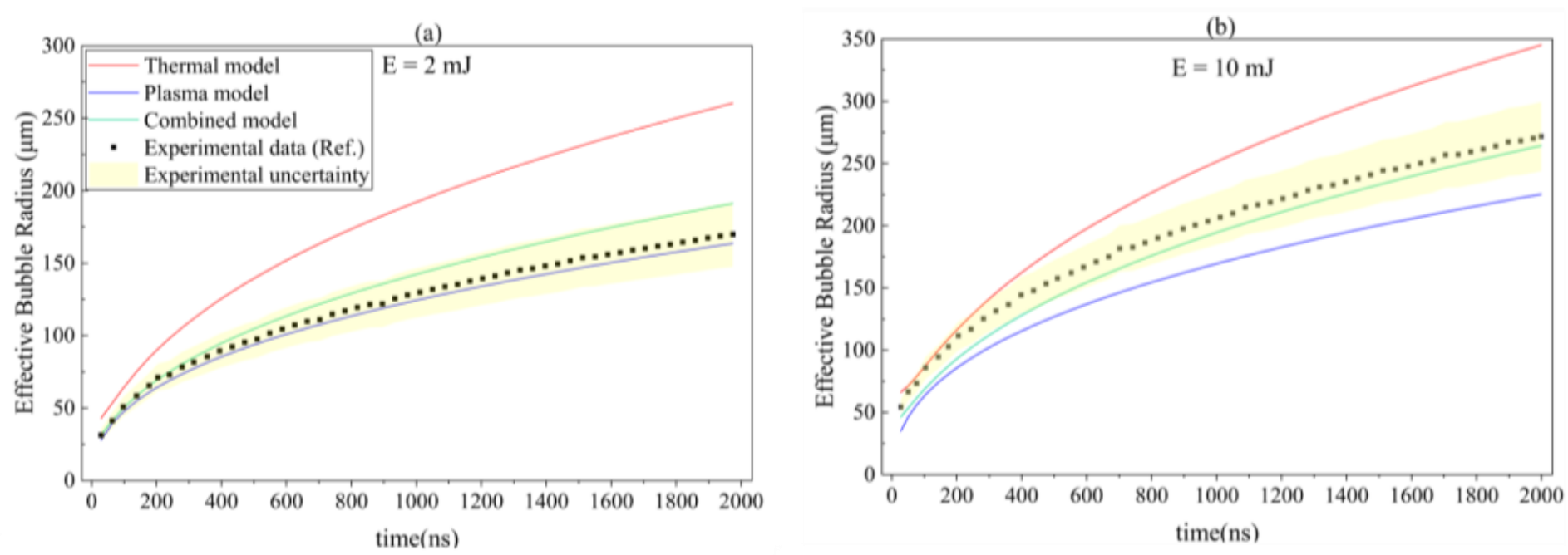

Figure 6. Representative validation case for early-stage bubble-inception modeling. Effective bubble-radius evolution is compared among thermal-only, plasma-mediated, and coupled plasma-thermal models at selected pulse energies of (a) 2 mJ and (b) 10 mJ. Experimental data and corresponding uncertainty bands are adapted from Jia et al. [76]. The comparison illustrates that different source-term assumptions lead to different bubble-growth predictions, whereas the coupled model provides a more physically constrained description of experimentally observed bubble dynamics. This example is included as an illustrative validation case for the early-stage laser-induced cavitation framework, not as a universal mechanism for all LIFT donor architectures. Model curves are adapted from Ref. [66].

For LIFT, the practical value of this early-stage framework lies not in the route comparison itself, but in the initial conditions it can provide. Thermal, plasma-mediated, or coupled models can estimate inception time, initial cavity size, internal pressure, residual thermal energy, bubble location, and geometric asymmetry. RP-type models can use these quantities to estimate the evolution of the equivalent radius. CFD/VOF or level-set models can use them to simulate confined bubble growth, free-surface deformation, jet formation, and droplet transfer. In this way, the framework serves as one possible mechanistic link between laser energy deposition and bubble-mediated printing, not as the chapter's universal explanation.

The framework is also useful because it separates questions that are often combined in empirical descriptions. One question is whether a bubble can be initiated at all under a given set of pulse conditions. A second is whether the newly formed cavity has enough pressure and energy to grow to a size that can deform the donor free surface. A third is whether the bubble position and shape promote forward jetting rather than lateral spreading or rupture. Thermal-only, plasma-mediated, and coupled routes address the first question in different ways, while RP and CFD models address the second and third questions more directly. This separation helps keep early laser physics, bubble growth, and jetting from being treated as a single adjustable threshold.

In applying such a framework to LIFT, the regime boundaries should be stated explicitly. A transparent, low-absorption liquid under intense short-pulse focusing may require optical breakdown physics. A strongly absorbing interlayer under nanosecond irradiation may be more accurately modeled as a confined thermal or thermoelastic source. A compliant dynamic release layer may require a mechanical blister model rather than a liquid cavitation model. The value of the three-route comparison is therefore diagnostic: it encourages the modeler to ask which source term is physically plausible for the donor architecture and laser parameters before simulating the visible bubble and jet. In this sense, it is one possible pathway for assigning early-stage conditions, not a replacement for the broader modeling hierarchy.

The comparison also clarifies why experiments should report intermediate observables. If a thermal model predicts a broad vapor region but imaging shows a narrow, elongated cavity, the assumed energy distribution may be incomplete. If a plasma-mediated impulse predicts early inception but underestimates later bubble size, residual thermal energy may be important. If an RP model reproduces radius but fails to predict jetting, confinement, and free-surface effects, it likely requires CFD. These diagnostic uses are regime-dependent and should be evaluated against measured bubble and jet dynamics.

## 10. Conclusions and Future Outlook

A bubble-aware view of LIFT leads to more practical design rules than fluence thresholds alone. The goal is not simply to exceed a transfer threshold, but to generate a bubble, vapor pocket, or blister-like impulse that is matched to the donor architecture, film thickness, material rheology, receiver gap, and desired deposit. Process development should therefore track intermediate observables such as bubble radius, bubble lifetime, jet speed, jet diameter, breakup state, impact behavior, and final deposit pattern.

Donor architecture is the first design lever. Absorbing layers should be selected not only for high optical absorption, but also for controlled thermal response, adhesion, compliance, and repeatability. Dynamic release layers and blister-actuated donors can protect sensitive inks, but they introduce additional mechanical and chemical parameters. Film thickness should be chosen with bubble-to-surface coupling in mind: thin films can favor efficient ejection and small transfer volumes, whereas thicker films may be useful for larger deposits or more viscous materials if sufficient impulse is available. Engineered interfaces, including porous or otherwise structured features, may offer one route to tune bubble-to-jet coupling or multi-jet ejection behavior, depending on geometry and operating conditions [67].

Laser parameters, material formulation, and receiver design should be treated as a coupled process window rather than as independent variables. Pulse energy, spot size, pulse duration, wavelength, and focusing determine the spatial and temporal loading of the donor, while viscosity, elasticity, particle or cell loading, receiver gap, wettability, and surface treatment determine how that loading is converted into jetting, breakup, impact, and final deposition. A useful process setting is therefore the one that produces the target bubble and jet observables while preserving material integrity.

This chapter reframed LIFT as a multiscale laser–liquid interaction process in which bubble dynamics bridge laser energy deposition and material ejection. Donor design, absorbing-layer response, laser loading, material properties, and receiver conditions jointly determine bubble inception, bubble growth, free-surface deformation, jet formation, droplet breakup, impact, and final pattern. Modeling approaches must therefore connect these intermediate states. Scaling analysis and energy-balance models identify dominant time and length scales; Rayleigh–Plesset-type models interpret equivalent-radius bubble growth; computational fluid dynamics, volume-of-fluid, and level-set simulations resolve confined interfaces and jetting; and thermal-fluid, plasma, thermoelastic, atomistic, or data-driven models can provide source terms, material inputs, or reduced-order process maps where appropriate.

Future progress will depend on experiments and models that validate intermediate observables rather than only final deposits. Benchmark datasets should include laser settings, donor geometry, material properties, time-resolved bubble and jet measurements, and deposition outcomes in a common format. These datasets would support predictive process maps for soft-material fabrication, where bubble dynamics, jet stability, and material integrity must be controlled together. For biomedical applications, future studies should connect bubble and jet observables not only to placement accuracy and deposit morphology, but also to cell viability, scaffold microstructure, and functional material performance. More broadly, this direction supports the long-term goal of building a physics-guided pathway toward high-resolution, high-throughput fabrication of multifunctional soft materials for tissue engineering, biosensing, wearable devices, and energy-related applications. With improved diagnostics, physically grounded simulations, and bubble-aware process maps, LIFT can move toward more predictive printing of functional, biological, particle-laden, and soft materials.

## Acknowledgements

The authors would like to acknowledge the funding support from the National Science Foundation (NSF) under CAREER Award No. CBET-2441995, "Laser-Induced Bubble Dynamics and Jet Formation in Complex Fluids," from the CBET Fluid Dynamics Program.

## Conflict of Interest

The authors declare no conflict of interest.